\documentclass[journal]{IEEEtran}

\usepackage{myStyleIEEE}

\usepackage[margin=0.65in]{geometry}
\usepackage{booktabs}
\usepackage{tabularx}
\usepackage{amsmath}
\usepackage{subcaption}
\usepackage{url}
\DeclareSIUnit \voltampere { VA } 
\DeclareSIUnit \var { var } 
\newcommand\mat[1]{\boldsymbol{#1}}
\IEEEoverridecommandlockouts

\begin{document}

\title{EMT and RMS Modeling of Thyristor Rectifiers for Stability Analysis of Converter-Based Systems}

\renewcommand{\theenumi}{\alph{enumi}}

\newcommand{\jovan}[1]{\textcolor{magenta}{$\xrightarrow[]{\text{J}}$ #1}}
\newcommand{\ognjen}[1]{\textcolor{pPurple}{$\xrightarrow[]{\text{O}}$ #1}}

\author{Ognjen~Stanojev,~\IEEEmembership{Member,~IEEE}
        Pol~Jane~Soneira,~\IEEEmembership{Member,~IEEE}
        G\"osta~Stomberg,~\IEEEmembership{Member,~IEEE}
        Mario~Schweizer,~\IEEEmembership{Senior Member,~IEEE}

\thanks{O. Stanojev, M. Schweizer are with ABB Corporate Research, Baden-D\"attwil,  Switzerland, emails:\{ognjen.stanojev, mario.schweizer\}@ch.abb.com.}
\thanks{P. J. Soneira, G. Stomberg are with ABB Corporate Research, Mannheim, Germany, emails:\{pol.jane-soneira, goesta.stomberg\}@de.abb.com.}
}

\maketitle
\IEEEpeerreviewmaketitle

\begin{abstract}
Thyristor rectifiers are a well-established and cost-effective solution for controlled high-power rectification, commonly used for hydrogen electrolysis and HVDC transmission. However, small-signal modeling and analysis of thyristor rectifiers remain challenging due to their line-commutated operation and nonlinear switching dynamics. This paper first revisits conventional RMS-based modeling of thyristor rectifiers and subsequently proposes a novel nonlinear state-space EMT model in the dq domain that can be linearized for small-signal analysis. The proposed model accurately captures all the relevant dynamic phenomena, including PLL dynamics, the commutation process, and switching delays. It is derived in polar coordinates, offering novel insights into the impact of the PLL and commutation angle on the thyristor rectifier dynamics. We verify the RMS and EMT models against a detailed switching model and demonstrate their applicability through small-signal stability analysis of a modified IEEE 39-bus test system that incorporates thyristor rectifier-interfaced hydrogen electrolyzers, synchronous generators, and grid-forming converters. 
\end{abstract}


\section{Introduction} \label{sec:intro}
Thyristor rectifiers are the most mature and prominent technology for high-power rectification, offering robustness, cost-effectiveness, and DC voltage controllability. These characteristics make thyristor rectifiers the preferred choice in applications such as HVDC transmission, large motor drives (e.g., in mining applications), and controlled rectification for electrochemical processes (e.g., hydrogen electrolysis). Nevertheless, analytical modeling and stability analysis of thyristor rectifiers remain challenging due to their line-commutated operation and the nonlinearities of the switching process.

The design and analysis of electrical systems that incorporate thyristor rectifiers often rely on computer-based time-domain simulation tools capable of accurately reproducing instantaneous voltage and current waveforms, including thyristor switching patterns. These simulations are suitable for the analysis of power quality~\cite{Yang2017} and large-signal events such as commutation failure~\cite{Gao2023}, short-circuit faults~\cite{Dong2017}, and operation under current limits~\cite{Mirsaeidi2019}. Although these simulations offer valuable insights, they are computationally intensive and cannot provide deeper system-theoretic insights. Therefore, analytical modeling for small-signal stability analysis is required to uncover the properties of operating points, identify critical operating conditions, and provide tuning guidelines~\cite{Wang2019}.

Small-signal stability analysis in power systems is commonly performed separately for Root Mean Square (RMS) and Electromagnetic Transient (EMT) models~\cite{Lara2024} to address different aspects of system behavior and to meet the different demands of grid design and grid operation tasks. RMS models, also known as quasi-static phasor models, simplify the system by representing voltages and currents as quasi-steady-state phasor quantities. This allows for computationally efficient analysis of slower dynamic phenomena such as interactions of outer control loops, frequency stability, etc. EMT models extend the validity of RMS models to also consider fast dynamic phenomena, such as network resonances, PLL transients, and interactions of device-level converter controls~\cite{Markovic2021}.

The RMS modeling of thyristor rectifiers is well established, with its formulation and applications extensively documented in classical references \cite{BRPelly,Kundur,JosArrillaga,BinWu}. These models assume a constant, balanced three-phase supply and use fundamental frequency relationships derived from the converter switching waveforms, thereby enabling tractable analysis and efficient simulation. However, RMS models often neglect commutation angle variations or rely on their quasi-steady-state approximations \cite{Karawita2009,Kwon2018}. As a result, the dynamics of the commutation process are not precisely characterized. Several studies have sought to extend the applicability of the RMS framework \cite{Osauskas2003,Atighechi2014,Ndiwulu2024}. For example, \cite{Osauskas2003} augments the formulation with filter dynamics, an AC-system representation, and PLL synchronization, while \cite{Atighechi2014,Ndiwulu2024} introduce additional control loops to broaden the modeling envelope. Nevertheless, the key assumptions remain, thus limiting the accuracy of the model. 

To improve the model quality while retaining analytical tractability, recent work has pivoted toward impedance-based models that capture multi-harmonic and interaction dynamics.
Within this line of work, small-signal impedance and admittance models of thyristor rectifiers (and inverters) have been developed primarily in the context of Line Commutated Converter-based HVDC (LCC-HVDC) systems. In \cite{10226300}, a general sequence-frame impedance formulation was proposed that incorporates the overlap angle via switching functions. Similarly, frequency-domain admittance models based on single-side modulated state functions were derived in \cite{10363678} to enable multi-harmonic linearization and provide a clearer view of harmonic interactions. Further advances include harmonic state-space formulations that employ multiple switching functions to model the variable commutation angle dynamics accurately \cite{10177884}. For hybrid HVDC systems, impedance-based stability analysis has been extended to multi-terminal cascaded configurations, where sensitivity analysis and impedance reshaping help mitigate oscillatory interactions \cite{10965539}. In parallel, the influence of control-link time delays on small-signal dynamics has been quantified in \cite{10114946}, showing that even modest delays can significantly affect stability margins. Although these contributions enhance the understanding of harmonic interactions, they remain predominantly frequency-domain-oriented, which limits readiness for time-domain simulation, analysis of nonlinear effects, and generalization beyond HVDC-specific contexts.

The need for high-fidelity EMT modeling of thyristor rectifiers for eigenvalue analysis and transient studies has driven the development of state-space formulations based on state averaging \cite{Yi2019,Lu2020,Dong2021,Chen2022}. In \cite{Yi2019}, a linearized state-space representation of thyristor rectifiers is derived using an infinite-series converter approach and subsequently transformed into the frequency domain to enable stability assessment through the generalized Nyquist criterion. The concept of motion equations has been employed in \cite{Lu2020} to develop an EMT state-space model for multi-infeed LCC-HVDC systems.  Building on these foundations, \cite{Dong2021} proposes a continuous-time state-space model that incorporates firing-angle control, DC-side dynamics, and commutation overlap for small-signal analysis. The work in \cite{Chen2022} focused on accurately modeling the commutation dynamics. More recently, \cite{Du2024} introduced a Linear Time-Periodic (LTP) framework to embed switching dynamics and address the nondifferentiable nature of thyristor commutation. Complementing these analytical approaches, \cite{Yang2025} presents an enhanced RMS model utilizing neural networks to capture EMT transients and mitigate commutation-failure risks.

Despite these advancements, several limitations persist in existing EMT modeling approaches for thyristor rectifiers. First, a systematic treatment of delays introduced by switching and PLL mechanisms is often absent, even though these delays can significantly shape dynamic performance during fast transients \cite{Yi2019,Dong2021,Du2024,Yang2025}. This gap also appears in earlier dq-frame treatments of thyristor-controlled devices such as thyristor controlled rectifiers and thyristor controlled static compensators, which prioritize fundamental-frequency dynamics over switching-induced timing effects \cite{Finotti2014,Perkins1999}. Second, LTP-based and neural-network-enhanced formulations introduce substantial complexity and computational burden, limiting practicality in large-scale system studies~\cite{Du2024,Yang2025}. Third, most models are tailored to LCC-HVDC and do not generalize readily to other high-power rectifier-based systems (e.g., hydrogen electrolyzers). In addition, many formulations assume balanced network variables and periodic operation, which constrains accuracy under asymmetrical operation or nonstationary grid conditions. Finally, most existing studies on thyristor rectifier dynamics have focused on relatively simple system configurations, such as HVDC systems fed by synchronous generators and voltage sources \cite{Karawita2009}, electrolyzer–generator interactions \cite{Ndiwulu2024}, and the IEEE 9-bus system \cite{Bidadfar2016}. While such system configurations provide valuable insights into local control interactions and rotor-angle stability, they do not capture the complexity of large interconnected grids where multiple thyristor-based loads interact with diverse generation sources. The absence of studies addressing large-scale power systems containing thyristor rectifiers is largely due to the reliance on impedance models, which are well-suited for frequency-domain analysis but impractical for large-scale networks with several generation and load types. 

In this paper, we revisit the RMS-based modeling of thyristor rectifiers and propose a novel EMT formulation suitable for large-scale stability studies. Both RMS and EMT models are presented to provide a complete perspective on the relevant characteristics of a small-signal model. We begin by examining existing RMS models, tracing their derivation, and highlighting the assumptions that limit their accuracy under fast transients. Building on this foundation, we extend the state-of-the-art in several directions. Firstly, we derive a continuous-time nonlinear state-space model in the dq domain that incorporates PLL dynamics, accurately represents the commutation process, and accounts for switching-related delays. The model is derived in polar coordinates, which offers novel insights into the impact of the PLL and commutation angle on the thyristor rectifier dynamics. Unlike previous approaches that relied on analytical linearization during model development, which introduced additional assumptions and potential inaccuracies, our method derives a nonlinear model to ensure higher fidelity. When required, linearization can be performed numerically, while the nonlinear formulation remains directly applicable for EMT simulations in time-domain studies. Secondly, although the proposed model employs a state-averaging technique, we demonstrate that its validity extends well beyond the switching frequency of the rectifier. This is confirmed through time-domain simulations and extensive frequency-domain analysis, which demonstrate close agreement with the behavior of a switching thyristor rectifier model. Finally, we conduct a large-scale system-level analysis using eigenvalue-based stability studies on a modified IEEE 39-bus network that integrates hydrogen electrolyzer loads, synchronous generators, and grid-forming converters. This configuration has been overlooked in the literature, despite its critical importance for future grids.

The remainder of the paper is structured as follows. Firstly, in Sec.~\ref{sec:RMS_model}, we briefly review the basics of small-signal analysis and present the RMS model. Subsequently, in Sec.~\ref{sec:EMT_model}, we rigorously derive a novel EMT model of thyristor rectifiers. The RMS and EMT models are validated both in frequency and time-domain in Sec.~\ref{sec:model_validation}. Finally, small-signal stability analysis of the standard IEEE 39-bus system containing electrolyzer loads interfaced via thyristor rectifiers is performed in Sec.~\ref{sec:results}.

\section{Modeling and Stability Analysis Preliminaries}\label{sec:RMS_model}
This section provides an overview of power system modeling using differential-algebraic equations for EMT and RMS simulations and outlines the adopted approach to small-signal analysis. We then introduce two conventional RMS models of thyristor rectifiers which conform to this framework. 

\subsection{RMS and EMT Power System Modeling}
A power system model represents a network of interconnected generation and load components in the form of explicit differential-algebraic equations. By grouping the system’s differential variables $\boldsymbol{x}$, algebraic variables $\boldsymbol{z}$, and inputs $\boldsymbol{u}$, the DAE system can be formalized as
\begin{subequations} \label{eq:DAEsys}
\begin{align} 
    \ddt{\boldsymbol{x}}&=\boldsymbol{f}\bigl(\boldsymbol{x},\,\boldsymbol{u},\,\boldsymbol{z}\bigr), \quad \boldsymbol{x}(0) = \boldsymbol{x}_\mathrm{ini}, \label{eq:diffDAE}\\
    0&=\boldsymbol{g}\bigl(\boldsymbol{x},\,\boldsymbol{u},\,\boldsymbol{z}\bigr)\label{eq:algDAE},
\end{align} 
\end{subequations} where $\boldsymbol{x}_\mathrm{ini}$ denotes the initial state of the differential variables.
The structure of the above DAE system is influenced by the modeling approach adopted. RMS models focus on system behavior at the fundamental frequency, representing network variables as phasors under the assumption of quasi-steady-state conditions. This simplifies the grid model to algebraic equations and is typically valid for analyzing phenomena up to around 20 Hz, such as electro-mechanical interactions of machines.
In contrast, EMT models retain full time-domain resolution (with averaged converter switching to enable linearization) and capture fast dynamics, including device-level converter control, machine flux behavior, and line transients. These models extend the validity in the frequency range up to 1~kHz or higher, resulting in detailed and higher-order DAEs.

\subsection{Small-Signal Stability Analysis}
The small-signal model of a power system can be obtained by linearizing \eqref{eq:DAEsys} around a desired equilibrium $\boldsymbol{\xi}_{\rm o}=(\boldsymbol{x}_{\rm o},\boldsymbol{u}_{\rm o},\boldsymbol{z}_{\rm o})$ where $\boldsymbol{f}\bigl(\boldsymbol{\xi}_{\rm o}\bigr)=0$ and $\boldsymbol{g}\bigl(\boldsymbol{\xi}_{\rm o}\bigr)=0$. Using these properties, the linearized DAE system reads as
\begin{subequations} \label{eq:linDAEsys}
\begin{align} 
    \ddt{\boldsymbol{\Delta{x}}}&=\mat{A}_{\rm xx}\boldsymbol{\Delta{x}}+\mat{B}_{\rm xu}\boldsymbol{\Delta{u}}+\mat{A}_{\rm xz}\boldsymbol{\Delta{z}}, \\\boldsymbol{\Delta{x}}(0) &= \boldsymbol{x}_\mathrm{ini} - \boldsymbol{x}_{\rm o},\\
    0&=\mat{A}_{\rm zx}\boldsymbol{\Delta{x}}+\mat{B}_{\rm zu}\boldsymbol{\Delta{u}}+\mat{A}_{\rm zz}\boldsymbol{\Delta{z}},
\end{align}
\end{subequations}
where the respective deviations from the desired operating point are represented by ${\boldsymbol{\Delta{x}}}=\boldsymbol{x}-\boldsymbol{x}_{\rm o},\,{\boldsymbol{\Delta{u}}}=\boldsymbol{u}-\boldsymbol{u}_{\rm o},\,{\boldsymbol{\Delta{z}}}=\boldsymbol{z}-\boldsymbol{z}_{\rm o}$. The matrices $\mat{A}_{\rm xx},\,\mat{B}_{\rm xu}$, and $\mat{A}_{\rm xz}$ denote the Jacobians of $\boldsymbol{f}$ with respect to $\boldsymbol{x},\,\boldsymbol{u},\,\boldsymbol{z}$, respectively, and the matrices $\mat{A}_{\rm zx},\,\mat{B}_{\rm zu}$, and $\mat{A}_{\rm zz}$ denote the Jacobians of $\boldsymbol{g}$ with respect to $\boldsymbol{x},\,\boldsymbol{u},\,\boldsymbol{z}$, respectively. The Jacobians are evaluated at $\boldsymbol{\xi}_{\rm o}$. We assume that DAE \eqref{eq:DAEsys} is of index-1. Then, $\boldsymbol{A}_{\rm zz}$ is regular and the linearized DAE system~\eqref{eq:linDAEsys} can be rewritten as 
\begin{align} \label{eq:ODEsys}
    \ddt{\boldsymbol{\Delta{x}}}=\mat{\tilde{A}}\boldsymbol{\Delta{x}}+\mat{\tilde{B}}\boldsymbol{\Delta{u}},\quad \boldsymbol{\Delta{x}}(0) = \boldsymbol{x}_\mathrm{ini} - \boldsymbol{x}_{\rm o}
\end{align} where $\mat{\tilde{A}} = \mat{A}_{\rm xx}-\mat{A}_{\rm xz}\mat{A}_{\rm zz}^{-1}\mat{A}_{\rm zx}$ and $\mat{\tilde{B}} =\mat{B}_{\rm xu}-\mat{A}_{\rm xz}\mat{A}_{\rm zz}^{-1}\mat{B}_{\rm zu}$ are the state-space matrices. 

To determine whether the system is small-signal stable, we analyze the eigenvalues ${\boldsymbol{\lambda}(\mat{\tilde{A}})}$ of the reduced state-space matrix. If all eigenvalues have negative real parts, then the power system is said to be small-signal stable at the equilibrium~$\boldsymbol{\xi}_{\rm o}$. 

\subsection{Thyristor Rectifier RMS Modeling}
The RMS modeling of thyristor rectifiers is well established in the power systems literature and is readily available in classical references \cite{BRPelly,Kundur}. 
In these works, the thyristor rectifier is modeled as a voltage source on the DC side and a current source on the AC side. The RMS behavior of a six-pulse thyristor rectifier, with input phase-to-ground voltage magnitude $V_\mathrm{m}$, firing angle $\alpha$, and output DC current $I_\mathrm{dc}$ is given by the following equation set:
\begin{subequations}\label{eq:RMS_eq_set1}
\begin{align}
	{V}_\mathrm{dc} &= \frac{3\sqrt{3}V_\mathrm{m}}{\pi}\cos{\alpha} - R_\mathrm{dc}I_\mathrm{dc}, \\
	I_\mathrm{m} &=  \frac{2\sqrt{3}}{\pi} I_\mathrm{dc},\quad \varphi=\alpha,
\end{align}
\end{subequations}
with $R_\mathrm{dc}=3\omega_\mathrm{g} L_\mathrm{c}/\pi$ modeling the commutation-induced voltage drop, and $L_\mathrm{c}$ denoting the commutation inductance. Thus, $V_\mathrm{dc}$ defines the voltage of the DC-side votlage source and $I_\mathrm{m}$ and $\varphi$ define the magnitude and phase displacement angle of the AC-side current source, respectively. This model captures the basic power conversion principles of a six-pulse thyristor rectifier. To more accurately capture the commutation-related effects, this model has been extended in \cite{JosArrillaga,BinWu}, resulting in
\begin{subequations}\label{eq:RMS_eq_set2}
\begin{align}
	V_\mathrm{dc} &= \frac{3\sqrt{3}V_\mathrm{m}}{\pi}\cos{\left(\alpha+\frac{\mu}{2}\right)}\cos{\left(\frac{\mu}{2}\right)},\\
    \mu &= \arccos\left( \cos(\alpha) - \frac{2\omega L_\mathrm{c}}{\sqrt{3}V_\mathrm{m}} I_\mathrm{dc} \right) - \alpha, \\
	I_\mathrm{m} &= \frac{2\sqrt{3}}{\pi} k_\mathrm{ic} I_\mathrm{dc}, \\
	\cos{\varphi} &= \frac{1}{k_\mathrm{ic}}\cos{\left(\alpha+\frac{\mu}{2}\right)}\cos{\left(\frac{\mu}{2}\right)},
\end{align}
\end{subequations}
where $\mu$ is the commutation angle and $k_\mathrm{ic}$ is defined as
\begin{equation*}
\resizebox{\columnwidth}{!}{$
\begin{aligned}
k_\mathrm{ic} &=
\frac{
\sqrt{
\bigl(\cos(2\alpha) - \cos\bigl(2(\alpha+\mu)\bigr)\bigr)^2
+ \bigl(2\mu + \sin(2\alpha) - \sin\bigl(2(\alpha+\mu)\bigr)\bigr)^2
}
}{
4\bigl(\cos(\alpha) - \cos(\alpha+\mu)\bigr)
}.
\end{aligned}
$}
\end{equation*}
Although \eqref{eq:RMS_eq_set2} provides greater accuracy, it does not introduce additional dynamics compared to \eqref{eq:RMS_eq_set1}; both models therefore exhibit identical dynamic behavior. Multipulse thyristor rectifier models, e.g., 12 pulse or 18 pulse,  adopt the same structure, differing only in the voltage and current coefficient values.

\subsection{RMS Model Assumptions and Limitations}\label{sec:ModelLimitations}
The previously presented model captures the dominant rectifier behavior at the fundamental frequency, making it suitable for RMS simulations. Its derivation is based on several simplifying assumptions, which limit its applicability to EMT simulations. Firstly, the input is assumed to be a balanced three-phase voltage of constant magnitude and frequency, whereas more comprehensive models would treat it as time-varying and adopt a dq domain representation. Secondly, the model omits the dynamics of the PLL, which is typically used to synchronize thyristor firing; such dynamics can introduce delays that lead to delayed or premature thyristor firing. Furthermore, a constant DC output current is assumed, although practical systems exhibit time-varying currents as a result of finite output inductance. Moreover, commutation inductance is considered only in terms of its low-frequency impact on the average DC voltage, neglecting its dynamic behavior at higher frequencies. Finally, AC-side currents are derived via Fourier series, based on assumed waveforms, rather than through direct, physics-based dynamic modeling, that would yield increased accuracy.

\section{EMT Modeling of Thyristor Rectifiers} \label{sec:EMT_model}
\begin{figure}[!b]
	\centering
	\includegraphics[scale=0.85]{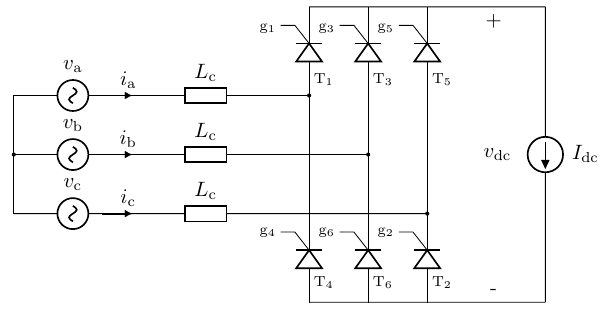}
	\caption{Configuration of a six-pulse thyristor rectifier supplied from a three-phase voltage source and feeding a constant current load.}
	\label{fig:SixPulseCircuit}
\end{figure}

In this section, we develop a more detailed nonlinear thyristor rectifier model suitable for EMT small-signal analysis by considering the thyristor differential equations during commutation and during conduction and averaging them over a switching period ($2\pi/p$ interval, where $p$ is the number of pulses of the thyristor rectifier). This model aims to overcome the limitations of the previously discussed RMS models. Without loss of generality, we consider the six-pulse rectifier topology depicted in Fig.~\ref{fig:SixPulseCircuit} for the subsequent derivations.

The DC voltage of a six-pulse thyristor rectifier consists of six identical segments within each fundamental cycle. That is, the DC voltage trajectory over each fundamental cycle is given by the concatenation of six equivalent segment trajectories. Thus, it is sufficient to model the DC voltage behavior over a single segment, i.e., by studying a single ${\pi}/{3}$ interval.
Throughout the \textit{commutation interval}, as the current is redirected from $T_6$ and $T_5$ to $T_6$ and $T_1$, the behavior of the system is governed by the following set of equations:
\begin{subequations}\label{eq:commutation_set}
\begin{align}
  L_\mathrm{c} \ddt{i_\mathrm{a}} &= v_\mathrm{a} - v_\mathrm{dc,p}, \qquad i_\mathrm{a} = i_\mathrm{a}, \\
  L_\mathrm{c} \ddt{i_\mathrm{b}} &= v_\mathrm{b} - v_\mathrm{dc,n}, \qquad i_\mathrm{b} = -I_\mathrm{dc}, \\
  L_\mathrm{c} \ddt{i_\mathrm{c}} &= v_\mathrm{c} - v_\mathrm{dc,p}, \qquad  i_\mathrm{c} = I_\mathrm{dc} - i_\mathrm{a},
\end{align}
\end{subequations}
where $v_\mathrm{dc,p}$ and $v_\mathrm{dc,n}$ denote the voltages of the positive and negative DC rails.
Similarly, during the \textit{conduction interval}, when $T_6$ and $T_1$ are conducting, the circuit model becomes:
\begin{subequations}\label{eq:conduction_set}
\begin{align}
  L_\mathrm{c} \ddt{i_\mathrm{a}} &= v_\mathrm{a} - v_\mathrm{dc,p}, \qquad i_\mathrm{a} = I_\mathrm{dc}, \\
  L_\mathrm{c} \ddt{i_\mathrm{b}} &= v_\mathrm{b} - v_\mathrm{dc,n}, \qquad i_\mathrm{b} = -I_\mathrm{dc}, \\
  v_\mathrm{c} &= v_\mathrm{c}, \qquad\qquad\quad\,\, i_\mathrm{c} = 0.
\end{align}
\end{subequations}
The equations outlined above are the basis for subsequent development of the EMT model. 

\subsection{DC Voltage Dynamics}
\begin{figure}[!b]
    \centering
    \includegraphics[scale=0.975]{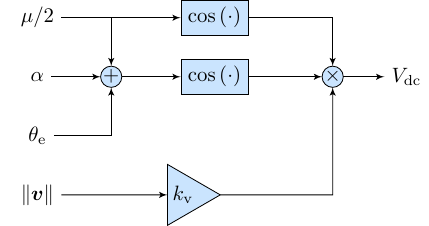}
    \caption{Control block diagram of the output DC voltage calculation.}
    \label{fig:Vdc_final}
\end{figure}

In order to capture both the commutation and conduction effects in a single continuous time expression, we average the conduction and commutation dynamics over a switching interval. To that end, the DC voltage $v_\mathrm{dc} = v_\mathrm{dc,p}-v_\mathrm{dc,n}$ can firstly be calculated for both the commutation (superscript com) and the conduction (superscript con) intervals, as: 
\begin{subequations}\label{eq:Vdc_con_com}
\begin{align}
    V_\mathrm{dc}^\mathrm{com} &= \frac{1}{2} v_\mathrm{a} + \frac{1}{2} v_\mathrm{c} - v_\mathrm{b} - \frac{3}{2}L_\mathrm{c}\ddt{I_\mathrm{dc}}, \\
    V_\mathrm{dc}^\mathrm{con} &= v_\mathrm{a} - v_\mathrm{b} - 2L_\mathrm{c}\ddt{I_\mathrm{dc}}.
\end{align}
\end{subequations}
Let us now consider a dq frame defined by an angle $\theta_\mathrm{r}$. A three-phase voltage vector $\boldsymbol{v}$ is described in the dq frame by a vector $\boldsymbol{v}=v_\mathrm{d}+jv_\mathrm{q}$ or in the polar form $\boldsymbol{v}=\|\boldsymbol{v}\|\exp{(\mathrm{j}\theta_\mathrm{e})}$.
The phase voltages in \eqref{eq:commutation_set} and \eqref{eq:conduction_set} supplying the rectifier can be converted to the dq domain by using the inverse Park transform:
\begin{subequations}\label{eq:abc2dq}
\begin{align}
v_\mathrm{a} &=  v_\mathrm{d} \cos \theta - v_\mathrm{q} \sin \theta, \\
v_\mathrm{b} &=  v_\mathrm{d} \cos \left(\theta - \frac{2\pi}{3}\right) - v_\mathrm{q} \sin \left(\theta - \frac{2\pi}{3}\right), \\
v_\mathrm{c} &=  v_\mathrm{d} \cos \left(\theta + \frac{2\pi}{3}\right) - v_\mathrm{q} \sin \left(\theta + \frac{2\pi}{3}\right).
\end{align}
\end{subequations}
Expressing the phase quantities in their dq forms, the DC voltage equations in the two intervals \eqref{eq:Vdc_con_com} become:
\begin{align*}
V_\mathrm{dc}^\mathrm{com} &= \frac{3}{2}\left( v_\mathrm{d} \cos(\theta + \frac{\pi}{3}) - v_\mathrm{q} \sin(\theta + \frac{\pi}{3}) \right) - \frac{3}{2}L_\mathrm{c}\ddt{I_\mathrm{dc}}, \\
V_\mathrm{dc}^\mathrm{con} &= \sqrt{3}\left( v_\mathrm{d} \cos(\theta + \frac{\pi}{6}) - v_\mathrm{q} \sin(\theta + \frac{\pi}{6}) \right) - {2}L_\mathrm{c}\ddt{I_\mathrm{dc}}.
\end{align*}
Unlike in the RMS model derivation, no assumptions are made here about the form of the three-phase AC source voltages. These expressions are hence exact, as no simplifications have been adopted thus far. Furthermore, it is interesting to observe that the commutation inductance appears on the DC side, i.e., multiplied by the derivative of the DC current.
The DC dynamics can now be averaged over a $\pi/3$ segment to obtain an EMT model:
\begin{equation*}
    {V}_\mathrm{dc} = \frac{3}{\pi}\left(\int_{-\frac{\pi}{3}+\alpha}^{-\frac{\pi}{3}+\alpha+\mu} V_\mathrm{dc}^\mathrm{com}(\theta) \, \mathrm{d}\theta 
+ \int_{-\frac{\pi}{3}+\alpha+\mu}^{\alpha} V_\mathrm{dc}^\mathrm{con}(\theta) \, \mathrm{d}\theta\right),
\end{equation*}
which, after evaluating the integrals and using $v_\mathrm{d}+jv_\mathrm{q}=\|\boldsymbol{v}\|\exp{(\mathrm{j}\theta_\mathrm{e})}$, yields the following DC voltage model
\begin{equation}\label{eq:Vdc_final}
{V}_\mathrm{dc} = k_\mathrm{v} \|\boldsymbol{v}\|  \cos\left( \frac{\mu}{2} \right) \cos\left( \alpha + \frac{\mu}{2} + \theta_\mathrm{e} \right) - \tilde{L}_\mathrm{c}(\mu)\ddt{I_\mathrm{dc}},
\end{equation}
where $k_\mathrm{v}={3\sqrt{3}}/{\pi}$ and $\tilde{L}_\mathrm{c}(\mu)=(2-{3\mu}/{2\pi})L_\mathrm{c}$.
Averaging the system dynamics inevitably introduces modeling errors, as it captures only the mean behavior rather than the full range of dynamic responses. Nevertheless, this approximation is essential for transforming the hybrid switched differential-algebraic equation system into a standard DAE formulation that is amenable to small-signal analysis. The extent of the modeling error introduced by this step is examined in the results section. Figure~\ref{fig:Vdc_final} illustrates the equation \eqref{eq:Vdc_final} in the form of a control block diagram. The variable inductance is included in the circuit representation, derived later in this section.

\subsection{Dynamics of the Commutation Process}
The DC voltage dynamics, previously derived in \eqref{eq:Vdc_final}, are a function of the commutation angle, denoted by $\mu$. As in the derivation of the RMS model, the commutation angle can be determined by analyzing the commutation process, which is governed by the following differential equation:
\begin{equation}\label{eq:ia_commutation}
2 L_\mathrm{c} \ddt{i_\mathrm{a}} = v_\mathrm{a} - v_\mathrm{c} + \underbrace{L_\mathrm{c} \ddt{I_\mathrm{dc}}}_{{\approx\,0}}.
\end{equation}
The differential equation is derived from the circuit in Fig.~\ref{fig:SixPulseCircuit} assuming commutation from thyristor $\mathrm{T}_5$ to thyristor $\mathrm{T}_1$, where phases a and c are shorted during the commutation interval until the current in $\mathrm{T}_1$, i.e., $i_\mathrm{a}$, reaches $I_\mathrm{dc}$.
The last term in \eqref{eq:ia_commutation} considers the effect of DC current variation during the commutation period, which is typically insignificant and can thus be neglected \cite{Chen2022}. It is important to note that there are no inherent mathematical challenges in solving the differential equation in its original form. The introduced simplification is rather adopted since the resulting expression depends on the difference between the DC current at the beginning and end of the commutation period. These quantities are not captured by the EMT model, as our approach relies on averaging. 

The phase voltages $v_\mathrm{a}$ and $v_\mathrm{c}$ can be expressed in the dq-domain as previously introduced in \eqref{eq:abc2dq} and their difference can be obtained using trigonometric identities, as
\begin{equation}
v_\mathrm{a} - v_\mathrm{c} = \sqrt{3} \left( v_\mathrm{d} \sin\left(\theta + \frac{\pi}{3}\right) + v_\mathrm{q} \cos\left(\theta + \frac{\pi}{3}\right) \right).
\end{equation}
Assuming $\omega = \ddt{\theta}$ is constant during commutation, the current $i_\mathrm{a}$ derivative can be written as
\begin{equation}
\ddt{i_\mathrm{a}} = \ddtheta{i_\mathrm{a}} \ddt{\theta} = \omega \ddtheta{i_\mathrm{a}}.
\end{equation}
Hence, the differential equation can be written as
\begin{equation}\label{eq:difi}
\ddtheta{i_\mathrm{a}} = \frac{\sqrt{3}}{2 \omega L_\mathrm{c}} \left( v_\mathrm{d} \sin\left(\theta + \frac{\pi}{3}\right) + v_\mathrm{q} \cos\left(\theta + \frac{\pi}{3}\right) \right),
\end{equation}
with the following initial and final conditions:
\begin{equation}
i_\mathrm{a}\left(\theta = -\frac{\pi}{3} + \alpha \right) = 0, \, i_\mathrm{a}\left(\theta = -\frac{\pi}{3} + \alpha + \mu \right) = I_\mathrm{dc}.
\end{equation}
The differential equation can be solved by integrating the expression in \eqref{eq:difi} from $\theta_0 = -\frac{\pi}{3} + \alpha$ to $\theta$, as follows
\begin{equation*}
i_\mathrm{a}(\theta) = \int_{\theta_0}^{\theta} \frac{\sqrt{3}}{2 \omega L_\mathrm{c}} \left( v_\mathrm{d} \sin\left(\phi + \frac{\pi}{3}\right) + v_\mathrm{q} \cos\left(\phi + \frac{\pi}{3}\right) \right) \mathrm{d}\phi,
\end{equation*}
which, after evaluating the integral, results in
\begin{equation}\label{eq:phase_a_exact}
i_\mathrm{a}(\theta) = \frac{\sqrt{3}\|\boldsymbol{v}\|}{2 \omega L_\mathrm{c}} \left( \cos(\alpha+\theta_\mathrm{e}) - \cos\left(\theta + \theta_\mathrm{e} + \frac{\pi}{3}\right) \right).
\end{equation}

\begin{figure}[]
    \centering
    \includegraphics[scale=0.9]{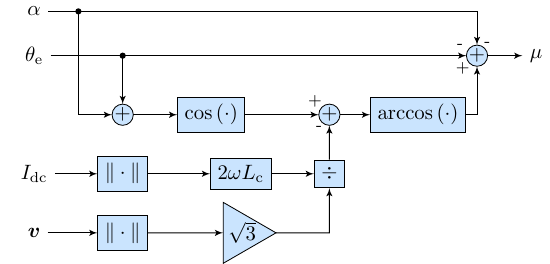}
    \caption{Control block diagram of the commutation angle calculation.}
    \label{fig:mu_diag}
\end{figure}

In order to determine the commutation angle $\mu$, let us impose the boundary condition for the current $i_\mathrm{a}(\theta)$:
\begin{equation}
I_\mathrm{dc} = \frac{\sqrt{3}\|\boldsymbol{v}\|}{2 \omega L_\mathrm{c}} \left( \cos(\alpha+\theta_\mathrm{e}) - \cos\left(\alpha + \mu + \theta_\mathrm{e} \right) \right).
\end{equation}
Solving the above equation in $\mu$, we obtain
\begin{equation}\label{eq:mu_final}
\mu = \arccos{\left(\cos{(\alpha+\theta_\mathrm{e})-\frac{2\omega L_\mathrm{c}}{\sqrt{3}\|\boldsymbol{v}\|}I_\mathrm{dc}}\right)}-\alpha-\theta_\mathrm{e}.
\end{equation}
This equation is illustrated in Fig.~\ref{fig:mu_diag} in the form of a control diagram. An absolute-value operator is introduced to enforce strictly unidirectional conduction within the thyristor rectifier.

\subsection{Input Current Dynamics}
\begin{figure}[!t]
    \centering
    \includegraphics[scale=0.975]{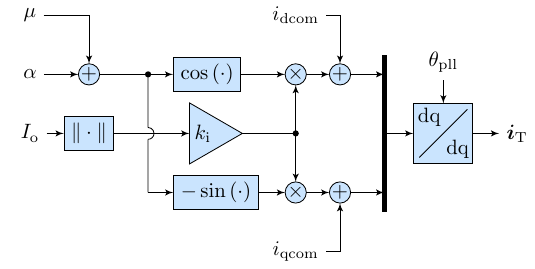}
    \caption{Control block diagram of the input current model.}
    \label{fig:input_current}
\end{figure}

\begin{figure}[!t]
    \centering
    \includegraphics[scale=0.975]{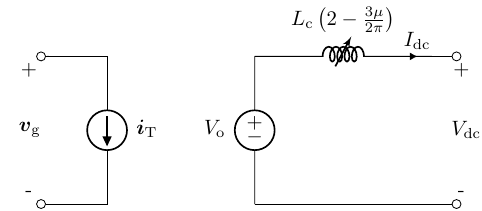}
    \caption{Circuit representation of the thyristor rectifier EMT model.}
    \label{fig:EMT_circuit}
\end{figure}

The average input current dynamics are derived following the same approach that is used to derive the DC voltage dynamics. As established at the beginning of this section, the phase currents during the \textit{commutation interval} are given by
\begin{equation}
i_\mathrm{a} =  i_\mathrm{a}, \qquad i_\mathrm{b} = -I_\mathrm{dc}, \qquad i_\mathrm{c} = I_\mathrm{dc} - i_\mathrm{a},
\end{equation}
which we transform to the dq domain through the Park transform, resulting in the following d- and q-components:
\begin{align}
i_\mathrm{d}^\mathrm{com}(\theta) &= \dfrac{2\sqrt{3}}{3} \left( i_\mathrm{a}(\theta) \sin\left(\theta + \frac{\pi}{3}\right) - I_\mathrm{dc} \sin \theta \right),\\
i_\mathrm{q}^\mathrm{com}(\theta) &= \dfrac{2\sqrt{3}}{3} \left( i_\mathrm{a}(\theta) \cos\left(\theta + \frac{\pi}{3}\right) - I_\mathrm{dc} \cos \theta \right).
\end{align}
On the other hand, the phase currents during the \textit{conduction interval} are given by
\begin{equation}
i_\mathrm{a} =  I_\mathrm{dc}, \qquad i_\mathrm{b} = -I_\mathrm{dc}, \qquad i_\mathrm{c} = 0,
\end{equation}
which we also transform to the dq domain, resulting in
\begin{align}
i_\mathrm{d}^\mathrm{con}(\theta) &= -\dfrac{2\sqrt{3}}{3} I_\mathrm{dc} \sin{(\theta-\frac{\pi}{3})},\\
i_\mathrm{q}^\mathrm{con}(\theta) &= -\dfrac{2\sqrt{3}}{3} I_\mathrm{dc} \cos{(\theta-\frac{\pi}{3})}.
\end{align}
The input current dynamics can now be averaged over a switching period to obtain the EMT model:
\begin{align*}
    {i}_\mathrm{d} &= \frac{3}{\pi}\left(\int_{-\frac{\pi}{3}+\alpha}^{-\frac{\pi}{3}+\alpha+\mu} i_\mathrm{d}^\mathrm{com}(\theta) \, \mathrm{d}\theta 
+ \int_{-\frac{\pi}{3}+\alpha+\mu}^{\alpha} i_\mathrm{d}^\mathrm{con}(\theta) \, \mathrm{d}\theta\right),\\
{i}_\mathrm{q} &= \frac{3}{\pi}\left(\int_{-\frac{\pi}{3}+\alpha}^{-\frac{\pi}{3}+\alpha+\mu} i_\mathrm{q}^\mathrm{com}(\theta) \, \mathrm{d}\theta 
+ \int_{-\frac{\pi}{3}+\alpha+\mu}^{\alpha} i_\mathrm{q}^\mathrm{con}(\theta) \, \mathrm{d}\theta\right).
\end{align*}
The final expressions depend on how the phase current $i_\mathrm{a}(\theta)$ during commutation is modeled. The expression in \eqref{eq:phase_a_exact} is accurate but complex. For this reason, we first employ a simplified form assuming that the current increases linearly during commutation. In this case, the phase current becomes
\begin{equation}
i_\mathrm{a}(\theta) = \frac{1}{\mu}I_\mathrm{dc}(\theta+\frac{\pi}{3}-\alpha).
\end{equation}
Under this assumption, the average current dynamics take the following form:
\begin{align}
    {i}_\mathrm{d} &= \dfrac{2\sqrt{3}}{3} I_\mathrm{dc}\cos{(\alpha+\frac{\mu}{2})},\\
    {i}_\mathrm{q} &=-\dfrac{2\sqrt{3}}{3} I_\mathrm{dc}\sin{(\alpha+\frac{\mu}{2})}.
\end{align}
Notice the resemblance between the equations above and the current approximation obtained in the RMS model~\eqref{eq:RMS_eq_set2}. However, if the exact expression in \eqref{eq:phase_a_exact} is adopted, we obtain:
\begin{align*}\label{eq:idq_final}
    {i}_\mathrm{d} &= \dfrac{2\sqrt{3}}{3} I_\mathrm{dc}\cos{(\alpha+\mu)} + {i}_\mathrm{dcom}, \\
    {i}_\mathrm{q} &=-\dfrac{2\sqrt{3}}{3} I_\mathrm{dc}\sin{(\alpha+\mu)} - {i}_\mathrm{qcom},
\end{align*}
where
\begin{align*}
{i}_\mathrm{dcom} &= \frac{3\|\boldsymbol{v}\|}{\pi\omega L_\mathrm{c}} \Bigg( 
2\sin\left(\alpha+\frac{\mu}{2}\right)\sin\left(\frac{\mu}{2}\right)\cos\left(\alpha+\theta_\mathrm{e}\right) \\
&\quad + \frac{\mu}{2}\sin\left(\theta_\mathrm{e}\right) 
- \frac{1}{2}\sin\left(\mu\right)\sin\left(2\alpha+\mu+\theta_\mathrm{e}\right) 
\Bigg), \\
{i}_\mathrm{qcom} &= \frac{3\|\boldsymbol{v}\|}{\pi\omega L_\mathrm{c}} \Bigg( 
2\cos\left(\alpha+\frac{\mu}{2}\right)\sin\left(\frac{\mu}{2}\right)\cos\left(\alpha+\theta_\mathrm{e}\right) \\
&\quad + \frac{\mu}{2}\cos\left(\theta_\mathrm{e}\right) 
+ \frac{1}{2}\sin\left(\mu\right)\cos\left(2\alpha+\mu+\theta_\mathrm{e}\right) 
\Bigg).
\end{align*}
The proposed input current model is depicted in Fig.~\ref{fig:input_current} as a control block diagram for clarity. 
With the behaviors on both the AC and DC sides now defined, an equivalent circuit can be constructed. The representation shown in Fig.~\ref{fig:EMT_circuit} illustrates that the thyristor rectifier EMT model is represented as a current sink on the AC side and as a voltage source on the DC side.
\subsection{PLL Dynamics}\label{subsec:PLLdynamics}
The PLL plays a critical role in rectifier operation by enabling the control system to accurately detect zero-crossing points and generate precise firing angles. Any error in the PLL angle results in delayed or premature thyristor firing, thereby affecting both the generated DC voltage and the input current. Naturally, commutation dynamics are also impacted.

\begin{figure}[!t]
    \centering
    \includegraphics[scale=0.975]{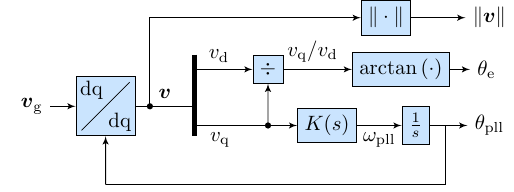}
    \caption{PLL control diagram.}
    \label{fig:pll_diagram}
\end{figure}

To adequately capture the dynamics of the PLL, the thyristor rectifier model should be implemented in a reference frame defined by the PLL angle $\theta_\mathrm{r}\equiv\theta_\mathrm{pll}$. A control block diagram of the considered PLL implementation is depicted in Fig.~\ref{fig:pll_diagram}. It can be observed that the angle $\theta_\mathrm{e} = \arctan(v_\mathrm{q}, v_\mathrm{d})$ appears consistently in the final expressions of the derived EMT model. Assuming the thyristor rectifier model is implemented in a reference frame defined by the PLL angle, the angle $\theta_\mathrm{e}$ represents the error in PLL angle estimation. By coupling the PLL error with the angle $\theta_\mathrm{e}$ appearing in the derived EMT model equations, the PLL dynamics are effectively incorporated into the rectifier EMT model. One can verify that such an approach leads to proper accounting of delayed and premature thyristor firing effects. Consequently, the following equations should be included in the EMT model to account for the PLL dynamics:
\begin{subequations}\label{eq:pll}
\begin{align}
    \boldsymbol{v} &= \mathcal{R}(\theta_\mathrm{pll})\boldsymbol{v}_\mathrm{g},\\
    \omega_\mathrm{pll} &= \omega_\mathrm{n} + K_\mathrm{pll,p} v_\mathrm{q} + \int_0^t K_\mathrm{pll,i} v_\mathrm{q} \mathrm{d}\tau,\\
    \ddt \theta_\mathrm{pll} &= \omega_\mathrm{pll}, \; \theta_\mathrm{pll}(0) = 0\\
    \theta_\mathrm{e} & = \arctan(v_\mathrm{q},v_\mathrm{d}),\\
    \|\boldsymbol{v}\| &= \sqrt{v_\mathrm{d}^2+v_\mathrm{q}^2},
\end{align}
\end{subequations}
where $\boldsymbol{v}_\mathrm{g}$ is the source voltage in the grid frame, $\omega_\mathrm{n}$ is the nominal angular frequency and $\mathcal{R}(\cdot)$ is the rotation matrix.

\subsection{Switching Delay Modeling}
Given that the six-pulse rectifier operates at a pulse frequency of $f_\mathrm{sw}=300\,\mathrm{Hz}$, high-frequency oscillations in the firing angle reference are not directly reflected in the output. To improve the accuracy of the model, we introduce a first-order low-pass filter with a cut-off frequency defined as $f_\mathrm{c} = {f_\mathrm{sw}}/{2}=150\,\mathrm{Hz}$ to prefilter the firing angle in the EMT model. The filtered firing angle is represented as
\begin{equation}\label{eq:lpf_alpha}
    \ddt{\alpha} = \omega_z \bigl(\alpha^\star - \alpha\bigr),\, \alpha(0)=\alpha^\star(0),
\end{equation}
where $\omega_z = 2\pi f_\mathrm{c}$ denotes the filter cut-off frequency, and $\alpha^\star$ is the reference firing angle typically determined by a controller.

\subsection{EMT Model Summary}
In summary, within the EMT model framework, the thyristor rectifier can be represented as a voltage source with a variable inductance on the DC side and a current sink on the AC side. This representation captures higher frequency behavior of the system, in contrast to the RMS-based modeling approach typically used for steady-state or low-frequency analysis. Besides the PLL dynamics~\eqref{eq:pll}, all equations pertaining to the EMT model are collected below:
\begin{subequations}\label{eq:EMT_equation_set}
\begin{align}
    {V}_\mathrm{dc} &= k_\mathrm{v} \|\boldsymbol{v}\|  \cos\left( \frac{\mu}{2} \right) \cos\left( \alpha + \frac{\mu}{2} + \theta_\mathrm{e} \right) - \ddt \tilde{L}_\mathrm{c}I_\mathrm{dc}, \\
    \mu &= \arccos{\hspace*{-0.5mm}\left(\hspace*{-0.5mm}\cos{(\alpha+\theta_\mathrm{e})-\frac{2\omega L_\mathrm{c}}{\sqrt{3}\|\boldsymbol{v}\|}I_\mathrm{dc}}\right)\hspace*{-1mm}}-\alpha-\theta_\mathrm{e}, \\
    {i}_\mathrm{d} &= \dfrac{2\sqrt{3}}{3} I_\mathrm{dc}\cos{(\alpha+\mu)} + {i}_\mathrm{dcom},\\
    {i}_\mathrm{q} &=-\dfrac{2\sqrt{3}}{3} I_\mathrm{dc}\sin{(\alpha+\mu)} + {i}_\mathrm{qcom}, \\
    \ddt {\alpha} &= \omega_z \bigl(\alpha^\star - \alpha\bigr),\; \alpha(0)=\alpha^\star(0),
\end{align}
\end{subequations}
with $i_\mathrm{dcom}$ and $i_\mathrm{qcom}$ defined in \eqref{eq:idq_final}, and $\boldsymbol{v}_\mathrm{g}$ and $I_\mathrm{dc}$ being independent inputs to the system. The proposed model is formulated for a six-pulse rectifier. Extensions to rectifiers with a higher number of pulses can be achieved in a straightforward manner by appropriately connecting multiple six-pulse rectifiers in series and/or parallel or by adjusting the corresponding coefficients in the model equations.

\section{Model Validation} \label{sec:model_validation}
In this section, we validate the reviewed RMS model \eqref{eq:RMS_eq_set1} and the proposed EMT model \eqref{eq:EMT_equation_set}  by comparing their dynamic behavior against the detailed switching model -- depicted in Fig.~\ref{fig:SixPulseCircuit} and with its switching operation described in~\cite{BRPelly}. The models are evaluated in both the time and frequency domains. Frequency-domain analysis is performed via impedance spectroscopy on both the DC and AC sides, measuring the DC output impedance and the AC input admittance. On the other hand, time-domain analysis focuses on the transient response to a load step change.
\begin{figure}[!t]
	\centering
	\includegraphics[scale=0.66]{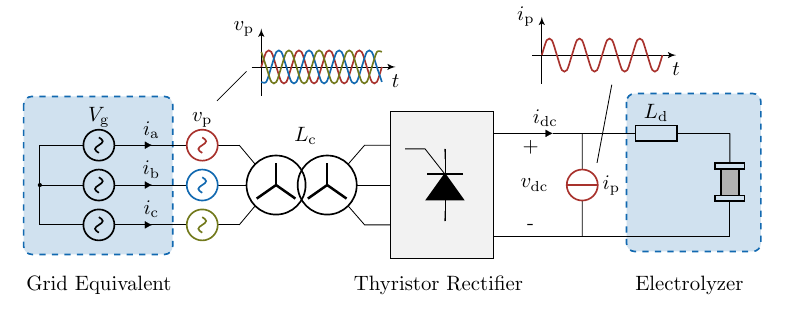}
	\caption{Impedance spectroscopy through load and source perturbations.}
	\label{fig:ImpedanceScanCircuit}
\end{figure}

\subsection{Time Domain Validation}
\begin{figure}[!b] 
	\centering
	\includegraphics[scale=1.125]{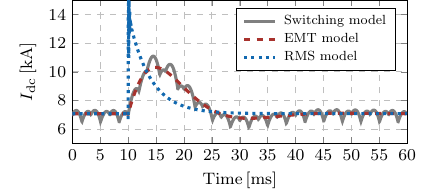}
	\caption{Time-domain response of the DC output current of the three considered models to a step change in the reverse voltage of the hydrogen electrolyzer.}
	\label{fig:TimeDomain_Validation}
\end{figure}

We first compare the time domain responses of the three above-mentioned models. The simulation results for the system in Fig.~\ref{fig:ImpedanceScanCircuit} are presented in Fig.~\ref{fig:TimeDomain_Validation}. The perturbation variables $v_\mathrm{p}=0$ and $i_\mathrm{p}=0$ are set to zero and the electrolyzer model is given in Sec.~\ref{sec:results} in \eqref{eq:electrolyzer}. At $10\,$ms simulation time, the reverse voltage of the electrolyzer is halved in a stepwise fashion, creating a short transient in the DC current. As can be seen from the figure, the predictions from the considered RMS and EMT models match well the switching model in steady-state. Nevertheless, during the transient, only the EMT model correctly predicts the behavior of the switching model. Therefore, model matching in higher frequencies requires more detailed analysis, which is performed subsequently using frequency domain approaches. 

\subsection{DC Output Impedance Comparison}
\begin{figure}[!t] 
	\centering
	\includegraphics[scale=1.125]{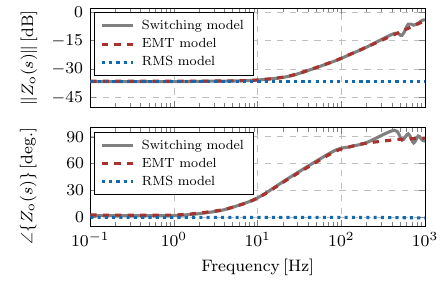}
	\caption{Comparison of DC output impedances of the RMS, EMT and full switching models.}
	\label{fig:DCOutputZCI}
\end{figure}
In this subsection, we verify how well the considered stability models (RMS and EMT) approximate the DC side dynamics by investigating the properties of the DC output impedance. The DC output impedance can be estimated using the circuit illustrated in Fig.~\ref{fig:ImpedanceScanCircuit}.
That is, load current perturbations $i_\mathrm{p}$ of various frequencies (in the frequency range of interest) are applied, while the electrolyzer load is disconnected and the source perturbations are not applied, i.e., $v_\mathrm{p}=0$. Subsequently, the output impedance is calculated as
\begin{equation*}
    Z_\mathrm{out}(\omega) =-\frac{v_\mathrm{dc}(\omega)}{i_\mathrm{dc}(\omega)},\qquad \omega\in[\omega_\mathrm{min},\omega_\mathrm{max}].
\end{equation*}

The output impedance scanning results are presented in Fig.~\ref{fig:DCOutputZCI}. Interestingly, the output impedance in the RMS model is constant and corresponds to $R_\mathrm{dc}$. This is an expected result, as the model includes only the low-frequency effects on the DC voltage. Higher frequency effects are well captured by the EMT model. The commutation inductance appears not only as a constant (resistive) component but also as an inductive element. The EMT model accurately captures the behavior of the full switching model, also at higher frequencies. The tiny mismatches at some frequencies between the EMT and the full model arise due to nonlinearities and switching harmonics.
\begin{figure*}[!t]
    \centering
    \includegraphics[width=\textwidth]{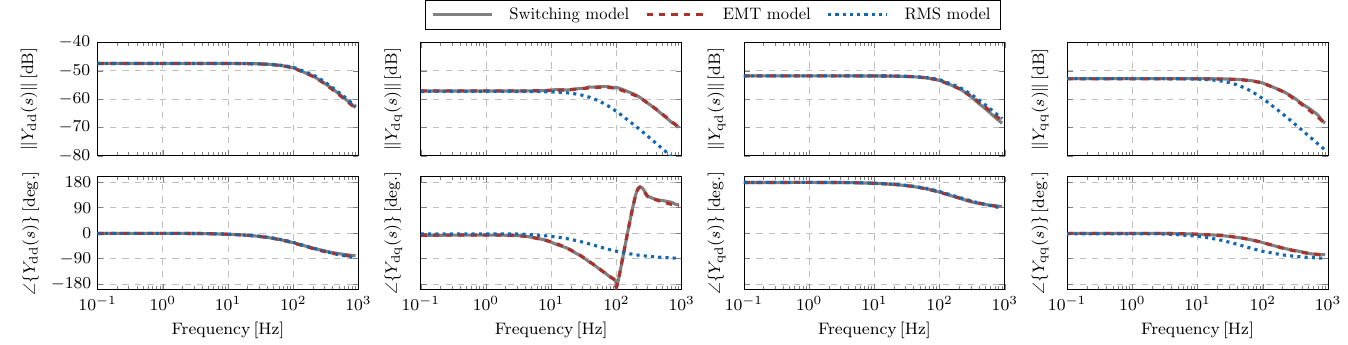}
    \caption{All four components of the input admittance, presented in both magnitude and phase, of the six-pulse rectifier with hydrogen electrolyzer load.}
    \label{fig:Y_input_admittance}
\end{figure*}

\subsection{AC Input Admittance Comparison}\label{sec:ACInputAdmCI}
Secondly, we verify how well the considered stability models approximate the AC side dynamics by investigating the properties of the AC input admittance.  AC input admittance is estimated using the circuit in Fig.~\ref{fig:ImpedanceScanCircuit} with $i_\mathrm{p}=0$ and the electrolyzer load connected to the DC terminal.

The objective of the frequency scanning procedure is to determine the {dq}-admittance of the system as observed from the point of perturbation. This is achieved by injecting a small voltage disturbance $v_\mathrm{p}$ at a specific frequency $\omega$ around the steady-state operating condition. By sweeping across a wide range of frequencies, the system's {dq}-admittance can be identified.
The {dq}-admittance consists of four components: $Y_\mathrm{dd}$, $Y_\mathrm{dq}$, $Y_\mathrm{qd}$, and $Y_\mathrm{qq}$, calculated as \cite{Ramakrishna2023}:
\begin{align}
    Y_\mathrm{dd}(\omega) = \frac{i_\mathrm{d}^\mathrm{dp}(\omega)}{v_\mathrm{d}^\mathrm{dp}(\omega)}, \quad
    Y_\mathrm{dq}(\omega) = \frac{i_\mathrm{d}^\mathrm{qp}(\omega)}{v_\mathrm{q}^\mathrm{qp}(\omega)},\\
    Y_\mathrm{qd}(\omega) = \frac{i_\mathrm{q}^\mathrm{dp}(\omega)}{v_\mathrm{d}^\mathrm{dp}(\omega)}, \quad
    Y_\mathrm{qq}(\omega) = \frac{i_\mathrm{q}^\mathrm{qp}(\omega)}{v_\mathrm{q}^\mathrm{qp}(\omega)}.
\end{align}

Here, the superscripts dp and qp denote the perturbations applied along the d and q axes, respectively, and $\omega\in[\omega_\mathrm{min},\omega_\mathrm{max}]$ represents the frequency at which the system is perturbed.
To accurately calculate dq-admittance, frequency scans must be performed independently for the d and q axes.

Figure~\ref{fig:Y_input_admittance} shows the measurements in the frequency-domain obtained using the proposed method. As illustrated in the plots, both the RMS and the EMT model reproduce the behavior of the full switching model accurately in the low-frequency region (up to approximately 10~Hz). However, the EMT model provides a substantially closer match to the full model, particularly in the $Y_{\mathrm{dq}}$ and $Y_{\mathrm{qq}}$ admittances. This improved matching is attributed to its higher fidelity, the inclusion of a PLL, and its detailed modeling of the commutation process.

\section{Small-Signal Stability Results} \label{sec:results}
This section presents small‑signal stability results for the IEEE 39‑bus test system comprising synchronous generators, grid‑forming converters, and electrolyzer loads interfaced through thyristor rectifiers. The objective is to assess the impact of thyristor rectifiers on overall system stability and to identify potential sources of instability. The section first describes the case study setup and system parameterization. It then employs modal analysis to identify the most critical system modes, followed by a bifurcation analysis of key thyristor rectifier and electrolyzer parameters that may precipitate system instability.

\subsection{Case Study Setup}
\begin{figure}[!b]
    \centering
    \includegraphics[scale=0.8]{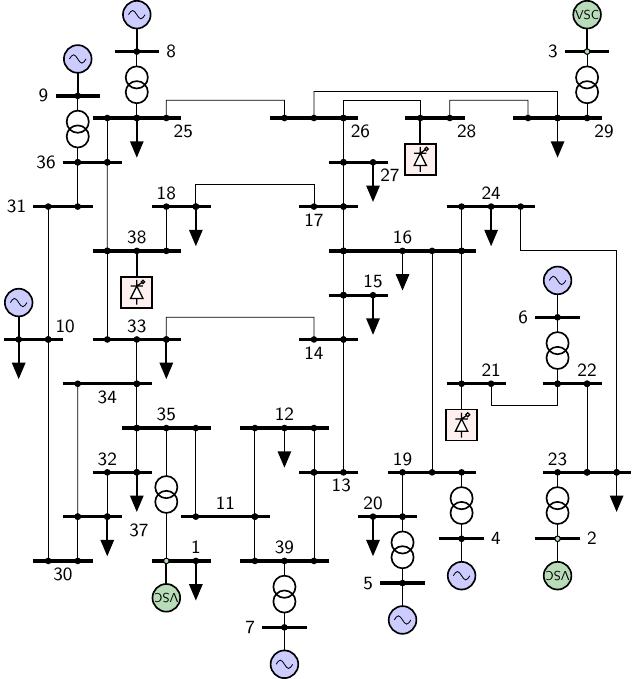}
    \caption{Single‑line diagram of the modified IEEE 39‑bus New England test system, illustrating the placement of VSC‑based generation and thyristor rectifier‑interfaced electrolyzer loads.}
    \label{fig:39busIEEE_diag}
\end{figure}

The IEEE 39‑bus New England test system considered in this paper is illustrated in Fig.~\ref{fig:39busIEEE_diag}. The original system~\cite{Athay1979} consists of ten conventional synchronous generators supplying a total load of approx. $6000\,\mathrm{MW}$. The system is modified to incorporate three electrolyzer loads, which are connected at buses~21, 28, and~38 with nominal power consumptions of $50\,\mathrm{MW}$,\, $100\,\mathrm{MW}$, and $200\,\mathrm{MW}$, respectively. To meet the additional electrolyzer demand, three renewable generation units interfaced through voltage source converters (VSCs) are added to buses~1, 2, and~3, with corresponding generation capacities of $50\,\mathrm{MW}$,\, $100\,\mathrm{MW}$, and $200\,\mathrm{MW}$. All VSCs are operated in grid‑forming mode. The relevant network, load, and synchronous generator parameters are adopted from the original reference~\cite{Athay1979}, while the modeling and parameterization of the electrolyzers and grid‑forming converters are described in the following. All simulations are performed in \textsc{PowerFactory}, which provides detailed models of synchronous machines, inverter‑interfaced generation, and transmission lines.\footnote{\url{https://www.digsilent.de/en/powerfactory.html}}

A Proton Exchange Membrane (PEM) hydrogen electrolyzer is considered in this work and is modeled using an equivalent electrical circuit. The electrolyzer cell voltage is described by the following set of equations~\cite{ElectrolyzerModeling2024}:
\begin{subequations}\label{eq:electrolyzer}
\begin{align}
V_\mathrm{cell} &= L_\mathrm{d}\ddt{I_\mathrm{cell}} + R_0 I_\mathrm{cell} + v_1 + V_\mathrm{rev},\\
C_1\ddt{v_1} &= I_\mathrm{cell} - \frac{v_1}{R_1},
\end{align}
\end{subequations}
where $L_\mathrm{d}$ denotes the DC‑side inductance, which is typically required for the stable operation of a thyristor rectifier. The parameter $R_0$ represents the membrane resistance, while $(R_1, C_1)$ define the parallel RC branch used to model the double‑layer capacitance effect. Finally, $V_\mathrm{rev}$ is the reversible voltage associated with the electrochemical reaction and corresponds to the minimum voltage required for hydrogen production.

The electrolyzer is supplied by a 12‑pulse thyristor rectifier, obtained by connecting in series two six-pulse rectifiers in~\eqref{eq:EMT_equation_set}. The rectifier DC voltage and current correspond directly to the electrolyzer cell voltage and current, that is, $V_\mathrm{dc} = V_\mathrm{cell}$ and $I_\mathrm{dc} = I_\mathrm{cell}$. To regulate hydrogen production, a proportional–integral controller is employed to adjust the firing angle of the thyristor rectifier, thereby tracking a desired DC current reference $I_\mathrm{dc}^\mathrm{ref}$. The parameter values of the three electrolyzers integrated into the IEEE 39‑bus system are summarized in Table~\ref{tab:electrolyzer_param} and are inspired by parameters in~\cite{Ndiwulu2024}.

A virtual synchronous machine model~\cite{DArco2014} is employed to represent grid‑forming converter behavior. As noted earlier, VSC‑based converters are connected at buses 1, 2, and 3, with rated capacities of $50\,\mathrm{MW}$, $100\,\mathrm{MW}$, and $200\,\mathrm{MW}$, respectively. The inertia and damping parameters are identical for all three converters, with an inertia constant of $5\,\mathrm{s}$ and a damping coefficient of $0.32$. Voltage magnitude regulation is implemented via a first‑order lag, consistent with standard virtual synchronous machine implementations.

\begin{table}[!t]
\renewcommand{\arraystretch}{1.2}
\caption{Electrolyzer parameters.}
\label{tab:electrolyzer_param}
\noindent
\centering
    \begin{minipage}{\linewidth}
    \begin{center}
    \scalebox{1}{%
        \begin{tabular}{ c || c | c | c | c | c | c}
            \toprule
            \textbf{Bus} & {Capacity} &$L_\mathrm{d}$ & $R_0$ & $R_1$ & $C_1$ & $V_\mathrm{rev}$ \\ 
            \cline{1-7}
             $38$ & \SI{200}{\mega\watt} & \SI{5}{\micro\henry} & \SI{0.2}{\milli\ohm} & \SI{0.7}{\milli\ohm} & \SI{53}{\farad} & \SI{100}{\volt}\\ 
             $28$ & \SI{100}{\mega\watt} & \SI{10}{\micro\henry} & \SI{0.4}{\milli\ohm} & \SI{1.5}{\milli\ohm} & \SI{25}{\farad} & \SI{100}{\volt}\\ 
             $21$ & \SI{50}{\mega\watt} & \SI{20}{\micro\henry} & \SI{0.8}{\milli\ohm} & \SI{3}{\milli\ohm} & \SI{10}{\farad} & \SI{100}{\volt}\\ 
            \arrayrulecolor{black}\bottomrule
        \end{tabular}
    }
        \end{center}
    \end{minipage}
\end{table}

\subsection{Modal Analysis}
In this subsection, we analyze the eigenvalues of the small‑signal model in \eqref{eq:linDAEsys} for the modified IEEE 39‑bus system at the nominal operating point described above. Since the system comprises a total of 253 modes, the analysis focuses on those eigenvalues located closest to the imaginary axis, as these are the most critical for stability assessment. A set of selected 21 critical eigenvalues is reported in Table~\ref{tab:eigenvalues}, expressed both as complex numbers in rectangular form and in terms of their corresponding damped frequencies $f_n = |\Im(\lambda)|/2\pi$ and damping ratios ${\zeta = -\Re{(\lambda)} / \sqrt{(\Re(\lambda)^2+\Im(\lambda)^2}}$. Furthermore, a participation‑factor analysis~\cite{perez-arriaga_selective_1982} is performed to determine the grid components that contribute to each mode, as shown in the last column of the table. We note that all of the analysis in this subsection and the next pertain solely to the EMT model. Stability issues with RMS thyristor rectifier models have not been observed in our study due to the absence of PLL and accurate current dynamics modeling. 

\begin{table}[!b]
\renewcommand{\arraystretch}{1.2}
\caption{Selected critical eigenvalues.}
\label{tab:eigenvalues}
\noindent
\centering
    \begin{minipage}{\linewidth}
    \renewcommand\footnoterule{\vspace*{-5pt}}
    \begin{center}
    \scalebox{0.9}{%
        \begin{tabular}{ c || c | c | c | c }
            \toprule
            \textbf{ID} &\textbf{Mode} & ${\zeta}$ & $f_n$ [Hz] & \textbf{Participating Units} \\ 
            \cline{1-5}
             $\lambda_1,\lambda_2$ & $-0.05\pm{j}0.03$ & $0.88$ & $0.004$ & $\mathrm{G}_{10},\mathrm{G}_9,\mathrm{VSM}_{3}$\\
            $\lambda_{3},\lambda_{4}$ & $-0.135\pm j0.437$ & 0.30 & 0.070 & $\mathrm{G}_9,\mathrm{VSM}_{3}$ \\
            $\lambda_5$  & $-0.188$ & 1.00 & 0 & $\mathrm{G}_{10},\mathrm{VSM}_{3}$ \\
            $\lambda_6,\lambda_7$  & $-0.198\pm j0.849$ & 0.23 & 0.135 & $\mathrm{G}_{10},\mathrm{G}_9$ \\
            $\lambda_8,\lambda_9$  & $-0.212\pm j0.451$ & 0.43 & 0.072 & $\mathrm{G}_{10},\mathrm{G}_4$ \\
            $\lambda_{10},\lambda_{11}$  & $-0.254\pm j0.519$ & 0.44 & 0.083 & $\mathrm{G}_{10},\mathrm{G}_4$ \\
            $\lambda_{12},\lambda_{13}$  & $-0.456\pm j6.074$ & 0.45 & 0.967 & $\mathrm{TR}_{28},\mathrm{TR}_{38},\mathrm{G}_9$ \\
            $\lambda_{14},\lambda_{15}$  & $-1.384\pm j0.350$ & 0.97 & 0.056 & $\mathrm{G}_{8},\mathrm{TR}_{28},\mathrm{VSM}_{3}$ \\
            $\lambda_{16},\lambda_{17}$  & $-4.833\pm j3.774$ & 0.79 & 0.601 & $\mathrm{G}_{10},\mathrm{TR}_{38}$ \\
            $\lambda_{18},\lambda_{19}$  & $-5.030\pm j3.424$ & 0.83 & 0.545 & $\mathrm{G}_{10},\mathrm{TR}_{28},\mathrm{VSM}_{3}$ \\
            $\lambda_{20},\lambda_{21}$  & $-5.162\pm j3.758$ & 0.81 & 0.598 & $\mathrm{G}_{10},\mathrm{TR}_{28}$,$\mathrm{TR}_{21}$ \\
            \arrayrulecolor{black}\bottomrule
        \end{tabular}
    }
        \end{center}
    \end{minipage}
\end{table}

The critical eigenvalues listed in Table~\ref{tab:eigenvalues} reveal the relative involvement of the synchronous generators (Gs), the virtual synchronous machine (VSM), and the thyristor rectifiers (TRs) in shaping the system's small-signal dynamics. The lowest-frequency electromechanical modes (e.g., $\lambda_{1,2}$ and $\lambda_{3,4}$) show dominant participation from the synchronous generators $\mathrm{G}_{10}$ and $\mathrm{G}_{9}$, together with the virtual synchronous machine $\mathrm{VSM}_{3}$. This indicates that the VSM is dynamically coupled to the classical rotor-angle oscillations and contributes to their damping. As the modal frequency increases, the influence of the synchronous machines remains present, but additional dynamics originating from the thyristor rectifiers become significant. In particular, modes such as $\lambda_{14,15}$ and $\lambda_{18,19}$ exhibit strong participation from $\mathrm{TR}_{28}$ and $\mathrm{TR}_{38}$, reflecting the higher sensitivity of rectifier current control loops to oscillations in the mid-frequency range. Several modes (e.g., $\lambda_{20,21}$) further highlight the combined interaction of synchronous generators, the VSM, and multiple electrolyzer loads.

\subsection{Bifurcation Analysis}
\begin{figure}[!b]
	\centering
	\begin{subfigure}{0.45\textwidth}
		\centering
		\scalebox{1.1}{\includegraphics[]{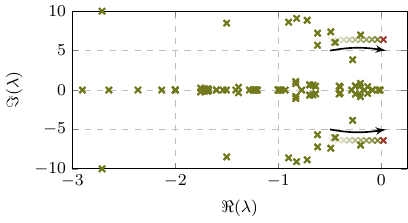}}
		\caption{\hspace*{-2.6em}}
		\label{fig:case3A_1}
		\vspace*{0.25cm} 
	\end{subfigure}
	\begin{subfigure}{0.45\textwidth}    
		\centering
		\scalebox{1.1}{\includegraphics[]{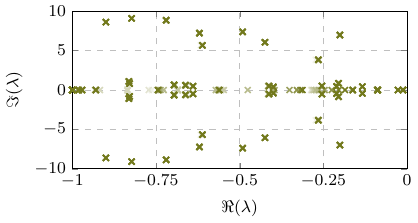}}
		\vspace*{-0.15cm}
		\caption{\hspace*{-2.6em}}
		\label{fig:case3A_2}     
	\end{subfigure}
	
	\caption{\label{fig:case3A} Movement of eigenvalues due to the changes in (a) PLL proportional gain; (b) current control bandwidth. }
\end{figure}

\begin{figure}[!t]
    \centering
    \includegraphics[scale=1.125]{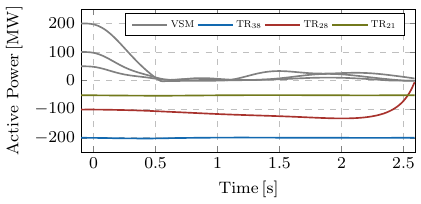}
    \caption{Trajectories of active power outputs of the three VSM converters and power consumption of the three thyristor rectifiers in the case of a PLL-induced instability.}
    \label{fig:Time_domain_instability}
\end{figure}

We further perform a bifurcation analysis to identify critical parameter values at which the system transitions from stable to unstable operation. The results indicate that these transitions are primarily driven by the tuning of the current controller and the PLL. Figure~\ref{fig:case3A_1} illustrates the evolution of the eigenvalue spectrum as the PLL proportional gain of the thyristor rectifier at bus 28 is reduced from $1$ to $0.01$, while Fig.~\ref{fig:case3A_2} depicts the effect of decreasing its current control bandwidth from $100$~Hz to $1$~Hz. In the former case, a bifurcation is observed, demonstrating that the system loses stability at the lowest examined PLL bandwidth. In particular, the eigenvalue pair $\lambda_{12,13}$ transitions toward and eventually crosses into the unstable region. In contrast, reducing the current controller bandwidth shifts the eigenvalues closer to the imaginary axis but does not induce instability within the considered parameter range. The evolution of the PLL-induced instability is presented in time domain in Fig.~\ref{fig:Time_domain_instability}. The system is initialized at the nominal operating point and the PLL proportional gain of the rectifer at bus 28 is set to $0.01$. As can be seen from the figure, the output power of the VSM units and the rectifier at bus 28 abruptly decrease to zero, leading to a system collapse.

\section{Conclusion} \label{sec:concl}
This work presented a nonlinear EMT model of a thyristor rectifier that achieves a balance between fidelity and computational tractability, addressing limitations of both RMS-based and time domain switched modeling approaches. By formulating rectifier dynamics in polar coordinates in the {dq} domain and incorporating PLL behavior, commutation overlap, and switching delays without restrictive assumptions, the model enables accurate EMT simulations as well as small-signal analysis of large-scale systems through numerical linearization. The validation results demonstrate close agreement with the full switching model. The eigenvalue-based stability analysis conducted on a modified IEEE 39-bus system further highlights the model's capability to reveal interactions among synchronous generators, grid-forming converters, and thyristor rectifiers. Furthermore, the study identified that PLL gains and current-control bandwidth of thyirstor rectifiers critically affect stability margins, with potential bifurcations emerging under poor PLL tuning. 

\bibliographystyle{IEEEtran}
\bibliography{bibliography}

\begin{thebibliography}{10}
\providecommand{\url}[1]{#1}
\csname url@samestyle\endcsname
\providecommand{\newblock}{\relax}
\providecommand{\bibinfo}[2]{#2}
\providecommand{\BIBentrySTDinterwordspacing}{\spaceskip=0pt\relax}
\providecommand{\BIBentryALTinterwordstretchfactor}{4}
\providecommand{\BIBentryALTinterwordspacing}{\spaceskip=\fontdimen2\font plus
\BIBentryALTinterwordstretchfactor\fontdimen3\font minus \fontdimen4\font\relax}
\providecommand{\BIBforeignlanguage}[2]{{%
\expandafter\ifx\csname l@#1\endcsname\relax
\typeout{** WARNING: IEEEtran.bst: No hyphenation pattern has been}%
\typeout{** loaded for the language `#1'. Using the pattern for}%
\typeout{** the default language instead.}%
\else
\language=\csname l@#1\endcsname
\fi
#2}}
\providecommand{\BIBdecl}{\relax}
\BIBdecl

\bibitem{Yang2017}
Y.~Yang, P.~Davari, F.~Zare, and F.~Blaabjerg, ``Enhanced phase-shifted current control for harmonic cancellation in three-phase multiple adjustable speed drive systems,'' \emph{IEEE Transactions on Power Delivery}, vol.~32, no.~2, pp. 996--1004, 2017.

\bibitem{Gao2023}
C.~Gao, J.~Yang, Z.~He, G.~Tang, J.~Zhang, T.~Li, and D.~He, ``Novel controllable-line-commutated converter for eliminating commutation failures of {LCC-HVDC} system,'' \emph{IEEE Transactions on Power Delivery}, vol.~38, no.~1, pp. 255--267, 2023.

\bibitem{Dong2017}
D.~Dong, Y.~Pan, R.~Lai, X.~Wu, and K.~Weeber, ``Active fault-current foldback control in thyristor rectifier for {DC} shipboard electrical system,'' \emph{IEEE Journal of Emerging and Selected Topics in Power Electronics}, vol.~5, no.~1, pp. 203--212, 2017.

\bibitem{Mirsaeidi2019}
S.~Mirsaeidi, X.~Dong, and D.~M. Said, ``A fault current limiting approach for commutation failure prevention in {LCC-HVDC} transmission systems,'' \emph{IEEE Transactions on Power Delivery}, vol.~34, no.~5, pp. 2018--2027, 2019.

\bibitem{Wang2019}
X.~Wang and F.~Blaabjerg, ``Harmonic stability in power electronic-based power systems: Concept, modeling, and analysis,'' \emph{IEEE Transactions on Smart Grid}, vol.~10, no.~3, pp. 2858--2870, 2019.

\bibitem{Lara2024}
J.~D. Lara, R.~Henriquez-Auba, D.~Ramasubramanian, S.~Dhople, D.~S. Callaway, and S.~Sanders, ``Revisiting power systems time-domain simulation methods and models,'' \emph{IEEE Transactions on Power Systems}, vol.~39, no.~2, pp. 2421--2437, 2024.

\bibitem{Markovic2021}
U.~Markovic, O.~Stanojev, P.~Aristidou, E.~Vrettos, D.~Callaway, and G.~Hug, ``Understanding small-signal stability of low-inertia systems,'' \emph{IEEE Transactions on Power Systems}, vol.~36, no.~5, pp. 3997--4017, 2021.

\bibitem{BRPelly}
B.~R. Pelly, \emph{\BIBforeignlanguage{eng}{Thyristor Phase-Controlled Converters and Cycloconverters}}, {First}~ed.\hskip 1em plus 0.5em minus 0.4em\relax John Wiley \& Sons, Inc., 1971.

\bibitem{Kundur}
P.~Kundur, \emph{\BIBforeignlanguage{eng}{Power System Stability and Control}}.\hskip 1em plus 0.5em minus 0.4em\relax McGraw-Hill, New York, 1994.

\bibitem{JosArrillaga}
J.~Arrillaga, \emph{\BIBforeignlanguage{eng}{High Voltage Direct Current Transmission}}, {Second}~ed.\hskip 1em plus 0.5em minus 0.4em\relax IET Power and Energy Series, 1998, vol.~29.

\bibitem{BinWu}
B.~Wu, \emph{\BIBforeignlanguage{eng}{High‐Power Converters and AC Drives}}, {First}~ed.\hskip 1em plus 0.5em minus 0.4em\relax John Wiley \& Sons, Inc., 2006.

\bibitem{Karawita2009}
C.~Karawita and U.~D. Annakkage, ``Multi-infeed {HVDC} interaction studies using small-signal stability assessment,'' \emph{IEEE Transactions on Power Delivery}, vol.~24, no.~2, pp. 910--918, 2009.

\bibitem{Kwon2018}
D.-H. Kwon, Y.-J. Kim, and S.-I. Moon, ``Modeling and analysis of an {LCC HVDC} system using dc voltage control to improve transient response and short-term power transfer capability,'' \emph{IEEE Transactions on Power Delivery}, vol.~33, no.~4, pp. 1922--1933, 2018.

\bibitem{Osauskas2003}
C.~Osauskas and A.~Wood, ``Small-signal dynamic modeling of {HVDC} systems,'' \emph{IEEE Transactions on Power Delivery}, vol.~18, no.~1, pp. 220--225, 2003.

\bibitem{Atighechi2014}
H.~Atighechi, S.~Chiniforoosh, J.~Jatskevich, A.~Davoudi, J.~A. Martinez, M.~O. Faruque, V.~Sood, M.~Saeedifard, J.~M. Cano, J.~Mahseredjian, D.~C. Aliprantis, and K.~Strunz, ``Dynamic average-value modeling of {CIGRE HVDC} benchmark system,'' \emph{IEEE Transactions on Power Delivery}, vol.~29, no.~5, pp. 2046--2054, 2014.

\bibitem{Ndiwulu2024}
G.~W. Ndiwulu, E.~V. Mayen, and E.~De~Jaeger, ``Dynamic interactions of a high-current thyristor-rectifier interfaced {PEM} electrolyzer with the angular dynamics of synchronous machines,'' in \emph{2024 IEEE International Conference on Environment and Electrical Engineering}, 2024, pp. 1--6.

\bibitem{10226300}
Y.~Chen, L.~Xu, A.~Egea-Àlvarez, and B.~Marshall, ``Accurate and general small-signal impedance model of lcc-hvdc in sequence frame,'' \emph{IEEE Transactions on Power Delivery}, vol.~38, no.~6, pp. 4226--4241, 2023.

\bibitem{10363678}
J.~Wang, Y.~Liu, C.~Fu, Z.~Wang, J.~Feng, H.~Li, Z.~Yu, Z.~Mo, and S.~Zhou, ``Frequency domain admittance model of line-commutated converter based on single-side modulated state function,'' \emph{IEEE Transactions on Power Delivery}, vol.~39, no.~2, pp. 845--858, 2024.

\bibitem{10177884}
T.~Liu, R.~Xu, Q.~Jiang, B.~Li, F.~Blaabjerg, and P.~Wang, ``Multiple switching functions based hss model of {LCC} considering variable commutation angle and harmonic couplings,'' \emph{IEEE Transactions on Power Delivery}, vol.~38, no.~6, pp. 3820--3833, 2023.

\bibitem{10965539}
R.~Xu, Q.~Jiang, B.~Li, Y.~Liu, T.~Liu, F.~Blaabjerg, and P.~Wang, ``Impedance based stability analysis of the multi-terminal cascaded hybrid hvdc system,'' \emph{IEEE Transactions on Power Delivery}, vol.~40, no.~3, pp. 1754--1768, 2025.

\bibitem{10114946}
C.~Guo, J.~Zhang, S.~Yang, and N.~Lv, ``Impact of time delay on the control link in small signal dynamics of {LCC-HVDC} system,'' \emph{IEEE Transactions on Power Delivery}, vol.~38, no.~5, pp. 3342--3355, 2023.

\bibitem{Yi2019}
Y.~Qi, H.~Zhao, S.~Fan, A.~M. Gole, H.~Ding, and I.~T. Fernando, ``Small signal frequency-domain model of a {LCC-HVDC} converter based on an infinite series-converter approach,'' \emph{IEEE Transactions on Power Delivery}, vol.~34, no.~1, pp. 95--106, 2019.

\bibitem{Lu2020}
J.~Lu, X.~Yuan, J.~Hu, M.~Zhang, and H.~Yuan, ``Motion equation modeling of {LCC-HVDC} stations for analyzing dc and ac network interactions,'' \emph{IEEE Transactions on Power Delivery}, vol.~35, no.~3, pp. 1563--1574, 2020.

\bibitem{Dong2021}
Y.~Dong, J.~Ma, S.~Wang, T.~Liu, X.~Chen, and H.~Huang, ``An accurate small signal dynamic model for {LCC-HVDC},'' \emph{IEEE Transactions on Applied Superconductivity}, vol.~31, no.~8, pp. 1--6, 2021.

\bibitem{Chen2022}
X.~Chen, J.~Ma, S.~Wang, T.~Liu, D.~Liu, and T.~Zhu, ``An accurate impedance model of line commutated converter with variable commutation overlap,'' \emph{IEEE Transactions on Power Delivery}, vol.~37, no.~1, pp. 562--572, 2022.

\bibitem{Du2024}
B.~Du, J.~Zhu, J.~Hu, Z.~Guo, S.~Ma, and J.~Guo, ``Small-signal modeling of {LCC-HVDC} considering switching dynamics based on the linear time-periodic {(LTP)} method,'' \emph{IEEE Transactions on Power Delivery}, vol.~39, no.~5, pp. 2715--2728, 2024.

\bibitem{Yang2025}
K.~Yang, X.~Wang, Q.~Zhang, G.~Geng, and Q.~Jiang, ``Dynamics enhanced quasi-steady-state model of {LCC-HVDC} systems based on neural network,'' \emph{IEEE Transactions on Power Delivery}, vol.~40, no.~4, pp. 2017--2028, 2025.

\bibitem{Finotti2014}
C.~Finotti and E.~Gaio, ``Continuous model in dq frame of thyristor controlled reactors for stability analysis of high power electrical systems,'' \emph{International Journal of Electrical Power \& Energy Systems}, vol.~63, pp. 836--845, 2014.

\bibitem{Perkins1999}
B.~Perkins and M.~Iravani, ``Dynamic modeling of high power static switching circuits in the dq-frame,'' \emph{IEEE Transactions on Power Systems}, vol.~14, no.~2, pp. 678--684, 1999.

\bibitem{Bidadfar2016}
A.~Bidadfar, H.-P. Nee, L.~Zhang, L.~Harnefors, S.~Namayantavana, M.~Abedi, M.~Karrari, and G.~B. Gharehpetian, ``Power system stability analysis using feedback control system modeling including {HVDC} transmission links,'' \emph{IEEE Transactions on Power Systems}, vol.~31, no.~1, pp. 116--124, 2016.

\bibitem{Ramakrishna2023}
R.~H. Ramakrishna, Z.~Miao, L.~Fan, and S.~Shah, ``{DQ} admittance extraction for inverter-based resources,'' in \emph{2023 IEEE Power \& Energy Society General Meeting (PESGM)}, 2023, pp. 1--5.

\bibitem{Athay1979}
T.~{Athay}, R.~{Podmore}, and S.~{Virmani}, ``A practical method for the direct analysis of transient stability,'' \emph{IEEE Transactions on Power Apparatus and Systems}, vol. PAS-98, no.~2, pp. 573--584, March 1979.

\bibitem{ElectrolyzerModeling2024}
M.~{Khalid Ratib}, K.~{M. Muttaqi}, M.~R. Islam, D.~Sutanto, and A.~P. Agalgaonkar, ``Electrical circuit modeling of proton exchange membrane electrolyzer: {The} state-of-the-art, current challenges, and recommendations,'' \emph{International Journal of Hydrogen Energy}, vol.~49, pp. 625--645, 2024.

\bibitem{DArco2014}
S.~D'Arco, J.~A. Suul, and O.~B. Fosso, ``Small-signal modelling and parametric sensitivity of a virtual synchronous machine,'' in \emph{2014 Power Systems Computation Conference}, 2014, pp. 1--9.

\bibitem{perez-arriaga_selective_1982}
I.~J. Perez-Arriaga, G.~C. Verghese, and F.~C. Schweppe, ``Selective {Modal} {Analysis} with {Applications} to {Electric} {Power} {Systems}, {PART} {I}: {Heuristic} {Introduction},'' \emph{IEEE Transactions on Power Apparatus and Systems}, vol. PAS-101, no.~9, pp. 3117--3125, 1982.

\end{thebibliography}



\end{document}